\documentclass[]{interact}

\usepackage{epstopdf}
\usepackage[caption=false]{subfig}
\usepackage{url,comment}

\usepackage[natbibapa,nodoi]{apacite}
\theoremstyle{plain}

\theoremstyle{definition}

\theoremstyle{remark}

\begin{document}


\title{Race and Religion in Online Abuse towards UK Politicians: Working Paper}

\author{
\name{Genevieve Gorrell\thanks{CONTACT Genevieve Gorrell. Email: g.gorrell@sheffield.ac.uk}, Mehmet E. Bakir, Mark A. Greenwood, Ian Roberts and Kalina Bontcheva}
\affil{The University of Sheffield, UK}
}

\maketitle

\begin{abstract}

Against a backdrop of tensions related to EU membership, we find levels of online abuse toward UK MPs reach a new high. Race and religion have become pressing topics globally, and in the UK this interacts with ``Brexit'' and the rise of social media to create a complex social climate in which much can be learned about evolving attitudes. In 8 million tweets by and to UK MPs in the first half of 2019, religious intolerance scandals in the UK's two main political parties attracted significant attention. Furthermore, high profile ethnic minority MPs started conversations on Twitter about race and religion, the responses to which provide a valuable source of insight. We found a significant presence for disturbing racial and religious abuse. We also explore metrics relating to abuse patterns, which may affect its impact. We find ``burstiness'' of abuse doesn't depend on race or gender, but individual factors may lead to politicians having very different experiences online.

\end{abstract}


\section{Introduction}

There is little doubt that the rise in social media usage is having far-reaching effects on society (e.g. \citeauthor{milan2015social}, \citeyear{milan2015social}), creating a need for vigilant commentary as the landscape shifts on almost a daily basis, in order to resist pernicious effects. Globally we have seen a rise in populism, fuelled by social media \citep{engesser2017populism}. Socially and politically, the impact of technological developments may be dramatic; for example \cite{allcott2017social} highlight the role of social media in the 2016 US presidential election. The United Kingdom, at the time of writing, are negotiating their exit from the European Union (``Brexit'') as a consequence of a referendum outcome that social media may have played a role in bringing about \citep{gorrell2019partisanship}.

In the spring of 2019, with Brexit forming a prominent but complex background issue to UK political dialogue, the subject of racism and religious tolerance arose in several notable ways. Immigration has been a key topic in the Brexit debate, centring on the UK's ability to control who enters. This, in conjunction with other global factors, gave oxygen to xenophobia with regards to migration of citizens \citep{rzepnikowska2019racism}.\footnote{\small{\url{https://www.independent.co.uk/news/science/brexit-prejudice-scientists-link-foreigners-immigrants-racism-xenophobia-leave-eu-a8078586.html}}} Dialogue is complex, however, with a strong presence of anti-racist voices, and the emergence of a polarized discussion in which identity politics may also be perceived as problematic \citep{fukuyama2018against}.

With the election in 2015 of Jeremy Corbyn as leader of the Labour Party (the opposition party in the UK's \textit{de facto} two party system) allegations of antisemitism in the party began to arise. The UK left wing has had a sympathy for the plight of Palestinians since the 1980s, but this hasn't previously been associated with antisemitism \citep{kolpinskaya2014book}. Additionally Boris Johnson, UK prime minister at the time of writing, has been found to have made several controversial comments about Islam,\footnote{\small{\url{https://www.theguardian.com/politics/2019/jul/15/boris-johnson-islam-muslim-world-centuries-behind-2007-essay}}} which, alongside incidents involving other Conservative politicians, has led to accusations of Islamophobia within the ruling Conservative Party.

In the same period, ethnic minority politicians received abuse that was in some ways unprecedented. We gathered a corpus of eight million tweets authored by UK MPs or sent in reply to them in the period of January to June 2019. We find that Brexit and Europe constitute the dominant topics. We also find that MP David Lammy, whose parents come from Guyana, received the most abusive tweets, in absolute terms, of any MP in that period. This means he received more abusive tweets than the then-prime minister, Theresa May, and more than the incoming prime minister, Boris Johnson, although Mr Johnson has long been a divisive figure who has always received high levels of abuse in our studies. Whilst the reason an MP receives abusive tweets is unlikely to depend only on skin colour, the matter nonetheless warrants investigation. Diane Abbott, whose parents are Jamaican, has also spoken up about the abuse she receives, which is widely regarded as unusual.\footnote{\small{\url{https://www.youtube.com/watch?v=1OZIqw5exdw}}}

These circumstances create an opportunity to explore the often complex and nuanced dialogue around racial and religious tolerance at a time when important transitions are being negotiated, both in terms of sovereignty and in terms of the feelings and opinions that are considered acceptable to express in civilized society. Abuse measured as volume or percentage fails to capture the experience certain MPs have described. Temporal variation may be a part of this (as, of course, is social context and the precise nature of the abuse). We proceed on the basis of the following research questions:

\begin{itemize}
    \item What abuse was received on Twitter by UK MPs in the first half of 2019? What topics were discussed and who was targeted? How are race and religion situated within this?
    \item What evidence do we find for racial and religious hate speech and discrimination in tweets to UK MPs? What individuals, events and themes emerge? How does this compare with other forms of discrimination in the corpus?
    \item How can we use temporal information to better understand the differing experience MPs have of receiving online abuse, in a way that goes beyond simple quantity?
\end{itemize}

\section{Related Work}
\label{sec:related}


Online fora have attracted much attention as a way of exploring political
dynamics \citep{nulty2016social,kaczmirek2013social,colleoni2014echo,weber2012mining,conover2011political,gonzalez2010emotional}. The effect of abuse and incivility in online political discussion has also been explored. \cite{vargo2017socioeconomic} find greater levels of online abuse in contexts characterized by poor socioeconomic prospects, low education levels and ``low partisan polarity'' (in other words, contexts characterized by a variety of opinion). With rising inequality and the opinion-dividing subject of Brexit, this suggests we might expect to encounter increased online abuse in the UK at the present time. \cite{gervais2015incivility} manipulates discourse in an online political forum, and finds that uncivil language from like-minded individuals escalates the incivility on the forum, whilst uncivil language from those in disagreement leads to withdrawal, suggesting a factor in the creation of uncivil echo chambers online. \cite{rusel2017bringing} finds provocative language use is more common than the vulgarity-centred abusive language on which this work focuses. With regards to online abuse toward politicians, \cite{theocharis2016bad} collected tweets centred around candidates for the European Parliament election in 2014 from Spain, Germany, the United Kingdom and France posted in the month surrounding the election, as a means to explore the role of incivility in politicians' motivations to engage, and thereby Twitter's political potential, at a time when it seemed more hopeful that the internet could be force for political good. (The following year, two of the same authors comment on the emerging realization that these hopes may have been misplaced \citep{tucker2017liberation}). \cite{ward2017turds} also make UK MPs their primary focus, exploring a two and a half month period running from late 2016 to early 2017. Our earlier work has explored Twitter abuse towards UK MPs in earlier time periods \citep{gorrell2018twits} and established trends spanning four years \citep{greenwood2019online} in which the topics producing strong responses have evolved in conjunction with the political climate. Given the developments online, concern is increasing about offline threats to public figures also. In this work we follow up on findings presented in the context of a BBC investigation\footnote{\small{\url{https://www.bbc.co.uk/news/uk-politics-49247808}}} into the increasing levels of threat and danger MPs are exposed to.


In the context of an increasing level of online toxicity, a surge of recent interest aims to detect abuse automatically. \cite{schmidt2017survey} provide a review of prior work
and methods, as do \cite{fortuna2018survey}. Whilst unintended bias has been the subject of much research in recent years with regards to making predictive systems that don't penalize minorities or perpetuate stereotypes, it has only just begun to be taken up within abuse classification \citep{park2018reducing}; unintended bias is a serious issue for sociological work such as that shared here. For that reason and others we adopt a rule-based approach here, as discussed below.


With regards to sexism, \cite{pew2017} find that women are twice as likely as men to receive sexist abuse online, and are also more likely to perceive online abuse as a serious problem. In terms of public figures, the subject of online abuse of women politicians and journalists has been taken up in a campaign by Amnesty International, e.g. \cite{stambolieva2017methodology}. A larger body of work has looked at hatred on social media more
generally \citep{perry2009cyberhate,coe2014online,cheng2015antisocial}. \cite{williams2015cyberhate} and \cite{burnap2015cyber} present work demonstrating the potential of Twitter for evidencing social models of online hate crime that could support prediction, as well as exploring how attitudes co-evolve with events
to determine their impact. \cite{silva2016analyzing} use natural language processing (NLP) to identify the groups targeted for hatred on Twitter and Whisper. \cite{farrell2019exploring} quantify and track misogyny online: we draw heavily on their lexica in our own abuse detection, as discussed below.

\section{Corpus and Methods}

In overview, our work investigates a large tweet collection on which a natural language processing has been performed to enable large scale quantitative analysis. In order to identify abusive language, the politicians it is targeted at, and the topics in the politician's original tweet that tend to trigger abusive replies, we use a set of NLP tools, combined into a semantic analysis pipeline. It includes, among other things, a component for MP and candidate recognition, which detects mentions of MPs. Topic detection finds mentions in the text of political topics (e.g. environment, immigration) and subtopics (e.g. fossil fuels). The list of topics was derived from the set of topics used to categorise documents on the gov.uk website\footnote{e.g. \small{\url{https://www.gov.uk/government/policies}}}, first
seeded manually and then extended semi-automatically to include
related terms and morphological variants using
TermRaider\footnote{\small{\url{https://gate.ac.uk/projects/arcomem/TermRaider.html}}}, resulting in a total of 1046 terms across 44 topics.
This methodology is presented in more detail by~\cite{Maynard17a}. A more detailed guide to the topics is also available online.\footnote{Warning; strong language and offensive slurs: \small{\url{https://gate-socmedia.group.shef.ac.uk/wp-content/uploads/2019/08/ICS-paper-supplementary-materials.docx}}}
We also perform hashtag tokenization, in order to find topic and abuse terms located within them. In this section we go into more detail regarding on the key steps of automatic abuse detection and collecting the tweets. We give an overview of the corpus thus obtained.

\subsection{Identifying Abusive Texts}

A rule-based approach was used to detect abusive language. An extensive vocabulary list of slurs, offensive words and potentially sensitive identity markers forms the basis of the approach.\footnote{\small{\url{https://gate-socmedia.group.shef.ac.uk/wp-content/uploads/2019/08/ICS-paper-supplementary-materials.docx}}} The slur list contained 1065 (UPDATE) abusive terms or short phrases in British and American English, comprising mostly an extensive collection of insults. Racist and homophobic slurs are included as well as various terms that denigrate a person's appearance or intelligence. We have included words from the work of \cite{farrell2019exploring}.

Offensive words such as the ``F'' word don't in and of themselves constitute abuse, but worsen abuse when found in conjunction with a slur, and become abusive when used with an identity term such as ``black'', ``Muslim'' or ``lesbian''. Furthermore, a sequence of these offensive words in practice is abusive, whereas a single such word is often not meant as abuse. NNN such words were used; examples include ``f**king'', ``sh*t'' and ``fat''. Similarly, identity words aren't abusive in and of themselves, but when used with a slur or offensive word, their presence allows us to type the abuse. Furthermore in assessing the degree of offensiveness, an attack against an identity group is usually considered more demeaning than an attack that doesn't reference such an aspect of who a person is. The ability to detect abuse where a sensitive term is coupled with an offensive word, rather than focusing only on slurs, means that more abuse is detected in this work than in previous work from our group.

Making the approach more precise as to target (whether the abuse is aimed at the politician being replied to or some third party) was achieved by rules based on pronoun co-occurrence. In the best case, a tight pronoun phrase such as ``you idiot'' or ``idiot like her'' is found, that can reliably be used to identify the target. Longer range pronoun phrases are less reliable but still useful. However, large numbers of insults contain no such adornment and are targeted at the tweet recipient, such as for example, simply, ``Idiot!''. Unless these are plurals, we count these.

Data from Kaggle's 2012 challenge, ``Detecting Insults in Social
Commentary''\footnote{\small{\url{https://www.kaggle.com/c/detecting-insults-in-social-commentary/data}}}, was used to evaluate the success of the approach. The training set was used to tune the terms included. On the test set, our approach was shown to have an accuracy of 80\%, and a precision/recall/F1 of 0.72/0.41/0.52. A higher precision is obtainable through use of tighter pattern matching regarding pronouns, but at the expense of recall. No particular bias is evident in the abuse detection; the missed phrases are usually more imaginative and lengthier ways of abusing, which are much harder to detect. In interpreting the findings presented here, it should be borne in mind that our method underestimates abuse.

Data-driven approaches have achieved widespread popularity through their typically superior results, but with the increasing real-world use of prediction systems, concerns have recently grown about their tendency to contain unwanted bias~\citep{hardt2016equality,bolukbasi2016man,caliskan2017semantics}. In the case of abuse detection, an example would be a system that learns from the training data that men receive more abuse, so on unseen data uses indicators of a male target, such as a male name, to classify a text as abusive. Subtler unwanted bias might include stereotyping a particular cultural group as more likely to be abusive, so labelling texts containing certain slang as abusive even when they aren't. An abundance of recent work seeks to address this problem, e.g.~\cite{dixon2018measuring,zhao2018gender}, but critics say more work is needed~\citep{elazar2018adversarial,pleiss2017fairness,gonen2019lipstick}. Our rule-based approach offers peace of mind and a fair performance as discussed in~\cite{greenwood2019online}. Additionally, a rule-based approach can be faster, which is important in processing large tweet collections such as the one used in this work. Furthermore, a rule-based approach can be easily extended with many rare words such as those crowdsourced in \url{hatebase.org}, which would require a very great deal of annotated training data to cover.

\subsection{Collecting Tweets}

The corpus were created by downloading tweets in real-time using Twitter's streaming API. The data collection focused on Twitter accounts of MPs, candidates, and official party accounts. We obtained a list of all current MPs\footnote{\small{\url{https://www.mpsontwitter.co.uk/list}}} (at that time) who had Twitter accounts.

We used the API to follow the accounts of all MPs over the period of interest. This means we collected all the tweets sent by each MP, any replies to those tweets, and any retweets either made by the MP or of the MPs own tweets. Note that this approach does not collect all tweets which an MP would see in their timeline as it does not include those in which they are just mentioned. We took this approach as the analysis results are more reliable  due to the fact that replies are directed at the politician who authored the tweet, and thus, any abusive language is more likely to be directed at them. Data were of a low enough volume not to be constrained by Twitter rate limits. Tables~\ref{tab:mps} and~\ref{tab:corpus} give statistics of the corpus thus collected, which contains a total of 153,878 MP-authored original tweets, 272,555 retweets and 73,707 replies. 7,531,817 replies to politicians were found, of which 3.92\% were reckoned to contain abuse.

\begin{table*}
\begin{center}
\resizebox{.9\columnwidth}{!}{%
  \begin{tabular}{l|l|l|l|l|l|l|l}
\textbf{Party} & \textbf{White/} & \textbf{White/} & \textbf{Minority/} & \textbf{Minority/} & \textbf{\% Fem} & \textbf{\% Min.} & \textbf{Total}\\
 & \textbf{Female} & \textbf{Male} & \textbf{Female} & \textbf{/Male} & \\
\hline
Conservative Party & 51 & 189 & 6 & 11 & 21.3 & 6.6 & 257\\
Labour Party & 94 & 109 & 18 & 11 & 46.3 & 12.5 & 232\\
Scottish National Party & 12 & 23 & 0 & 0 & 34.3 & 0.0 & 35\\
Liberal Democrats & 3 & 7 & 1 & 0 & 30.0 & 9.1 & 11\\
The Independent Group & 7 & 3 & 0 & 1 & 70.0 & 9.1 & 11\\
Democratic Unionist Party & 1 & 7 & 0 & 0 & 12.5 & 0.0 & 8\\
Independent & 0 & 7 & 0 & 0 & 0.0 & 0.0 & 7\\
Sinn Fein & 4 & 3 & 0 & 0 & 57.1 & 0.0 & 7\\
Plaid Cymru & 1 & 3 & 0 & 0 & 25.0 & 0.0 & 4\\
Green Party & 1 & 0 & 0 & 0 & 100.0 & 0.0 & 1\\
\hline
\textbf{Total} & \textbf{174} & \textbf{351} & \textbf{25} & \textbf{23} & \textbf{34.7} & \textbf{8.4} & \textbf{573}\\
 \end{tabular}
 }
\caption{MP statistics}
\label{tab:mps}
\end{center}
\end{table*}

\begin{table*}
\begin{center}
\resizebox{.85\columnwidth}{!}{%
  \begin{tabular}{l|l|l|l|l|l|l|l}
\textbf{Date} & \textbf{Orig} & \textbf{Retw. By} & \textbf{Repl. By} & \textbf{Repl. To} & \textbf{Abusive} & \textbf{\% Abusive} & \textbf{Sig Inc}\\
\hline
Jan & 25,131 & 43,758 & 12,391 & 980,413 & 31,650 & 3.23 & N/A\\
Feb & 24,133 & 43,002 & 11,396 & 1,098,061 & 38,933 & 3.55 & ***\\
Mar & 31,393 & 55,271 & 16,082 & 1,595,854 & 58,656 & 3.68 & ***\\
Apr & 23,009 & 42,064 & 11,906 & 1,165,899 & 45,204 & 3.88 & ***\\
May & 21,693 & 39,812 & 9,825 & 1,156,620 & 46,502 & 4.02 & ***\\
Jun & 27,822 & 47,763 & 11,670 & 1,503,795 & 72,915 & 4.85 & ***\\
 \end{tabular}
 }
\caption{Corpus statistics}
\label{tab:corpus}
\end{center}
\end{table*}

\section{Findings}

In this section we present findings under headings pertaining to each of the research questions in turn. We begin with an overall view of the corpus, that allows us to place racism and religious intolerance in their context, in terms of events and individuals, and quantify their importance. Next we investigate religious and racial hatred in more depth, in the context of other forms of discrimination. Finally we consider how abuse is experienced temporally by MPs and the impact it may therefore have.

\subsection{RQ1: Who Received Abuse, how much, and on what Topics?}

\begin{figure}
  \centering
\includegraphics[width=0.85\textwidth]{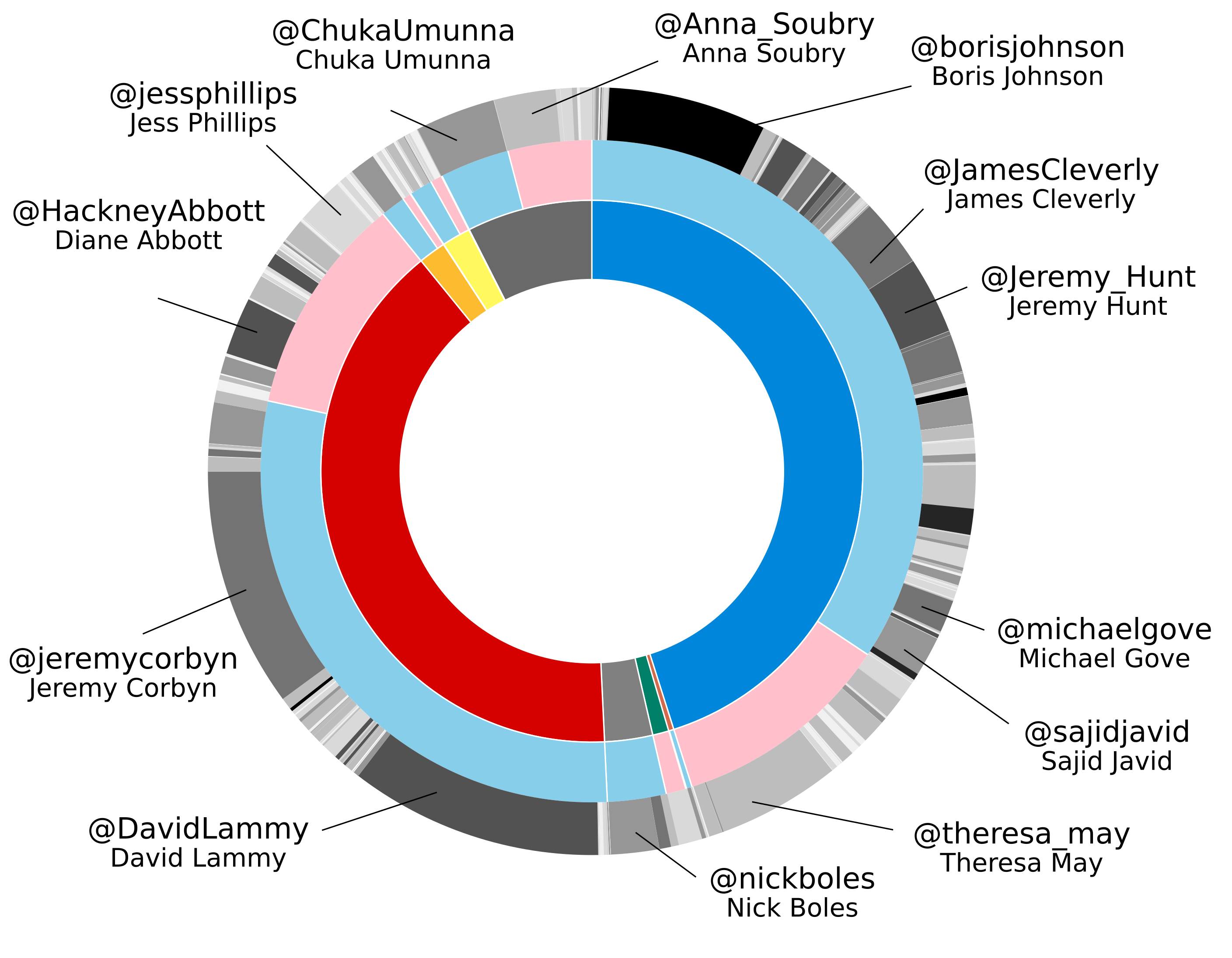}
  \caption{Who received the most abuse?}
   \label{fig:sunburst}
\end{figure}

Figure~\ref{fig:sunburst} presents the abuse received by the MPs during the period studied. Wider segments indicate a greater volume of abuse received in absolute terms, and darker segments indicate more abuse as a percentage of all replies received. Therefore a wide, light segment such as that representing outgoing prime minister Theresa May indicates that Mrs May received a lot of attention on Twitter, but that it didn't tend to be especially abusive quantitatively speaking. The ten most abused MPs by volume are annotated. Of those, the ethnic minority and mixed race MPs are Sajid Javid, David Lammy, Diane Abbott, Chuka Umunna and James Cleverly. It is evident at a glance that most of the abuse is received by a small number of high profile individuals, with most MPs receiving smaller volumes. An interactive version of the display, in which other MPs can be viewed and a fixed width version can also be viewed, is available online.\footnote{Withheld for anonymity.}

Labour MP David Lammy received the most abuse in absolute terms, receiving 11\% of all abuse found in the corpus. Jeremy Corbyn also received a high volume of abuse (10\% of all abuse found). In terms of abuse as a percentage of replies received, Boris Johnson was the high-profile politician with the highest percentage (8.39\%). Other politicians may receive a higher percentage due to low overall reply volumes. It is typical for Boris Johnson to receive consistently high percentages of abuse. Theresa May, Prime Minister at the time of data collection, received the highest volume of abuse of any female politician. However as a percentage, the level of abuse she receives is not unusually high (3.72\%). Diane Abbott received the second highest volume of abuse of any female politician, as well as the higher percentage of 6.35\%. Theresa May received 19\% of the total abuse sent to women MPs. Diane Abbott received 9\%. The ethnic minority MP receiving the highest volume of abuse was David Lammy, as can be inferred from his receiving the highest volume of abuse of any MP. His percentage of abuse received was 6.16\%, which is similar to the percentage received by Diane Abbott (6.35\%). David Lammy received almost four times as much abuse by volume as James Cleverly, who was the ethnic minority MP to receive the second most abuse by volume.

Some MPs do receive very high percentages of abuse, but where this is part of a low volume of replies the result may not be statistically significant, or it may be of less interest where it results from minor local events of less relevance to society. MPs with the highest percentages of abuse, where the result is statistically significant and the overall abuse volume is greater than 2,000 tweets are Boris Johnson (8.39\%), Ross Thomson (7.09\%), Daniel Kawczynski (6.77\%), Diane Abbott (6.35\%) and David Lammy (6.16\%).

\subsubsection{Topics}


\begin{figure}
  \centering
\includegraphics[width=0.95\textwidth]{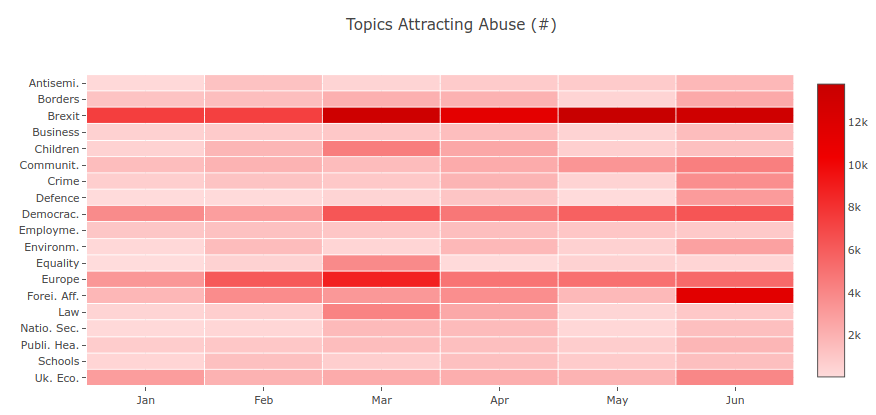}
  \caption{Topics attracting abusive replies: heat map}
   \label{fig:heatmap}
\end{figure}

The dominant topics of the period were ``Brexit'' (73,000 mentions by MPs), ``democracy'' (43,000 mentions) and ``Europe'' (41,000 mentions). ``Foreign affairs'' and ``community and society'' follow with 38,000 mentions. In addition to ongoing discussion of the exit arrangements, the European Parliament election occurred during this period, and was seen as an opportunity to demonstrate and clarify national feeling with regards to Europe despite the expectation of forfeiting representation in the European Parliament following exit. The will of the people, as ascertained through the democratic process, is a key concept with regards to Brexit.

In figure~\ref{fig:heatmap}, topics attracting over 5000 abusive replies across the six-month period are presented in the form of a heat map. This representation enables us to get a sense of the comparative abusiveness of many topics across the period that would be hard to see on a timeline. The data presented is on a per-month basis, and shows the absolute number of abusive replies that topic attracted in MP tweets. Note that precise values depend on the coverage of the topic lexicon. However, the dominant topics of the period are clearly illustrated. Generally speaking we can see that Brexit was a hot topic, and that feelings ran high in March and June. March is striking for the emergence of ``law and the justice system'' and ''equality, rights and citizenship'' as particularly abusive topics. This arose mainly from the case of Shamima Begum. Law/justice also frequently arises in the context of Brexit: March was the original month when the UK was planned to leave the EU, so feelings relating to legality ran high around that time.

In June, US president Donald Trump visited the UK, so it is unsurprising that ``foreign affairs'' appeared as an abusive topic in that month. Furthermore Mr Trump insulted (Muslim) London mayor Sadiq Khan, which led to supportive comments from among others David Lammy, who then came under fire (though Islam also appeared in abusive tweets in a variety of other contexts). Other prominent abusive topics were ``defence and the armed forces'' and ``national security'': Jeremy Corbyn is frequently abused by being called anti-military and a terrorist sympathizer. David Lammy is also called a terrorist sympathizer. Antisemitism appears, but Islamophobia does not, as it attracted far fewer abusive (and indeed non-abusive) remarks (c. 1,500 as opposed to over 5,000 abusive).

\begin{figure}
  \centering
\includegraphics[width=0.95\textwidth]{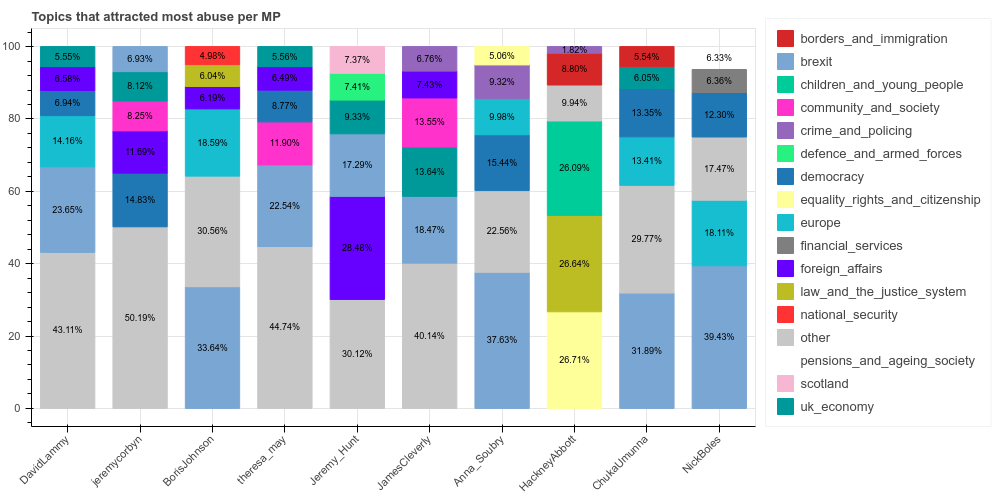}
  \caption{Topics attracting abusive replies: top MPs}
   \label{fig:mpbars}
\end{figure}

Figure~\ref{fig:mpbars} shows the topics in MP tweets that drew the most abuse, by volume, in reply, for the ten individuals receiving the most abuse in absolute terms. The five most abusive topics for each invididual are shown. These topics have a ``long tail''; many little-mentioned topics appear. For this reason, there is a substantial ``other'' category showing the extent of material that isn't shown, and would clutter the display to do so. Again, the focus on Brexit is evident. For two MPs, ``community and society'' appears as the topic attracting the most abuse. James Cleverly attracts significantly more abuse for his tweets on this topic than is typical for the topic (10.6\%, where 4.4\% is typical, p\textless0.001, Fisher's Exact test). Jeremy Corbyn attracted abuse for his comments on the subject of democracy; 6.0\%, where 3.6\% is typical for the topic (p\textless0.001). This is perhaps because he has been vocal in criticizing the process around Brexit negotiations and the election of a new leader by the Conservative Party. The most striking case of an MP attracting abuse for very different topics to usual, however, is Diane Abbott, who drew abuse for her comments on the subject of ``equality, rights and citizenship'' and ``law and the justice system''. This arises from her expressing support for Shamima Begum, a young woman who left the UK to join ISIL and was subsequently denied repatriation despite mortal risk to her unborn child. Again, these topics aren't Ms Abbott's usual topics; she tweets more on the subject of ``crime and policing'' and ``community and society''. However, her tweets on the subject of ``equality, rights and citizenship'' and ``law and the justice system'' drew 15.4\% and 15.3\% abusive replies respectively, where 8.1\% and 6.4\% respectively are typical (p\textless0.001 in both cases).

\begin{figure}
  \centering
\includegraphics[width=0.95\textwidth]{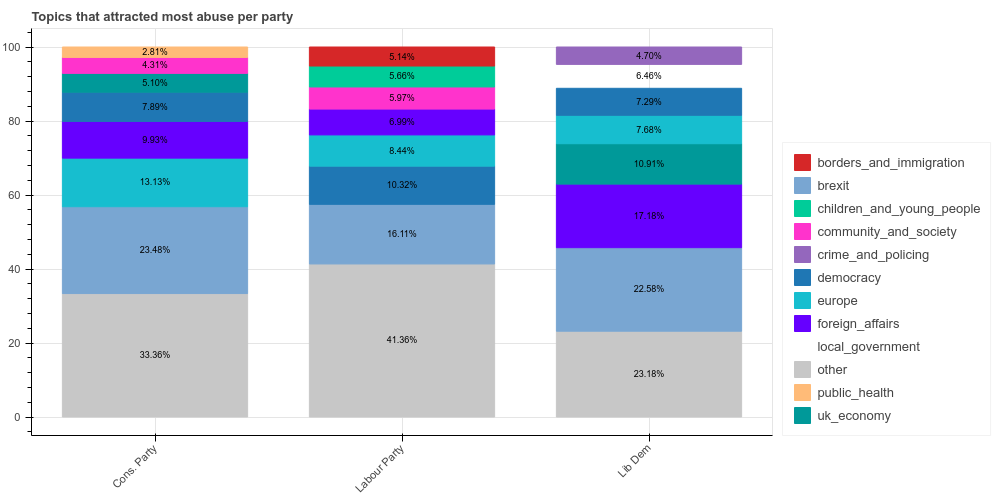}
  \caption{Topics attracting abusive replies by party}
   \label{fig:partybars}
\end{figure}

Figure~\ref{fig:partybars} gives the same topic display but divided along party lines for three parties, and in this case showing the seven most abusive topics for each party. In all cases, the Brexit/Europe focus is evident. Beyond that, we see some differences however. The UK economy appears more prominently for the Liberal Democrats. The Liberal Democrats were the minor partner in a coalition government with the Conservative Party during the recovery period from the 2007-2008 financial crisis--the austerity policy was controversial, and forms a theme in these tweets. For both the Conservative and Labour parties, however, ``community and society'' (a topic that subsumes Islam and Judaism) appears prominently as the fourth topic, which it doesn't for the Liberal Democrats. This reflects the party religious hatred scandals, though explicit discussion of antisemitism and Islamophobia was insufficient to appear separately in the graph. (A comment about ``Judaism'' for example would count toward ``community and society'' but not ``antisemitism'', resulting in higher counts for the broader discussion topic, in which other religions and sexual orientations are also pooled.)

\subsection{RQ2: What Hate Speech and Discrimination was Found?}

We found 15,500 politically abusive tweets directed at MPs (e.g. ``libtard'', ``effing Tory'' etc.) and almost 10,000 sexist tweets. We found 5,500 racist tweets directed at MPs. Homophobic, antisemitic and Islamophobic tweets were much fewer in number, comprising just a few hundred. Racism and sexism are quite broadly defined. We chose to include for example ``stupid man'' and ``stupid woman'' as sexist abuse, a decision based in part on the observation that ``stupid woman'' occurs approaching three times as often in tweets to MPs as ``stupid man'', despite women comprising only a third of MPs. Political abuse might be expected in a corpus of tweets directed at MPs. Racist abuse includes milder abuse against national groups that aren't marginalized, e.g. ``bloody French'', and borderline racist terms such as ``spiv'' and ``shyster'', as well as very extreme racist abuse. However broadly speaking, the quantities found are indicative of the extent of the phenomena.

We found, as in previous work, that men received more incivility across all types than women, receiving 2.24\% abusive replies ($\sigma$=1.76) vs. 1.56 ($\sigma$=1.14), a difference significant at p\textless0.001 using the Mann Whitney test. Women MPs received significantly more sexist abuse (\%, $\sigma$= vs. \%, $\sigma$= , p\textless0.001). The highest volume of sexist abuse was received by Theresa May, with over 1,000 sexist abusive replies. Emily Thornberry was found to have the highest percentage of abusive replies, at 0.52\%. David Lammy received the second highest volume of sexist abusive replies (recall that we include e.g. ``stupid man'' and ``stupid woman''), but at 1.4\% of his replies received, ranks him 12th behind 11 women for percentage of sexist abuse received. Diane Abbott (0.49\%) and Naz Shah (0.37\%) receive the second and third highest percentages of sexist abuse respectively. Men received more political abuse (\%, $\sigma$= vs. \%, $\sigma$= , p\textless0.05), a result that echoes~\cite{pew2017}.

Ethnic minority MPs received significantly more racist abuse (\%, $\sigma$= vs. \%, $\sigma$= , p\textless0.05). They also received significantly more Islamophobic abuse (\%, $\sigma$= vs. \%, $\sigma$= , p\textless0.001) and antisemitic abuse (\%, $\sigma$= vs. \%, $\sigma$= , p\textless0.05). As noted above, racism is quite broadly defined in the corpus, so strong racist abuse was considered separately. Strength of abuse was gauged by the number of components of the abuse. For example, ``stupid white idiot'' would be counted as having three components, and therefore considered stronger than ``bloody French'', having two components. When considering only strong racist abuse, i.e. that which had at least three components, ethnic minority MPs received again significantly more (\%, $\sigma$= vs. \%, $\sigma$= , p\textless0.05). Ethnic minority MPs also received significantly more abuse of all types having at least four components (\%, $\sigma$= vs. \%, $\sigma$= , p\textless0.05).

As a percentage and in terms of volume, David Lammy received the most racist abuse, with 0.22\% racist replies and a total number in excess of 1,000. Sajid Javid received 0.18\% racist abuse (), and Diane Abbott received 0.13\% (). Interestingly, the MP receiving the second most racist replies in terms of volume was Boris Johnson, with almost 300 racist abusive replies (although as a percentage this was only \%). On inspection, this is because people have used the words ``spiv'' and ``shyster'' to abuse him. These words have some historical association with Judaism, and Boris Johnson does have some Jewish ancestry, as well as ancestry of various kinds. It is unclear whether these abusive tweets were intended to have racist connotations in addition to their main meaning (calling into question Mr Johnson's honesty)--racism can be deliberately subtle--or whether no racism was intended. Sajid Javid and James Cleverly received in the region of 200 racist abusive replies, making them third and fourth for overall volume.

Conservative MPs received more abuse than Labour Party MPs, with MPs of other parties too small in number to produce reliable observations. Labour MPs received 1.70\% abusive replies ($\sigma$=1.46) compared with 2.30\% for Conservative MPs ($\sigma$=1.70), a difference significant at p\textless0.001. Where multiple factors are in play, it is hard to disentangle the key elements, but previous work~\citep{gorrell2018twits} has attempted a combined model with modest success, and found gender to be the only reliable predictor of abuse. The greater abuse received by Conservative MPs may arise from their being in power, and from the gender balance of their MPs. Conservative MPs also received more racist abuse despite their lower ethnic minority representation, perhaps because Conservatives draw ``spiv'' and ``shyster'' more than Labour politicians, and because hostility against the English (for example by the Scottish) is counted (\%, $\sigma$= vs. \%, $\sigma$= , p\textless0.01), less Islamophobic abuse (\%, $\sigma$= vs. \%, $\sigma$= , p\textless0.05) and more political abuse (\%, $\sigma$= vs. \%, $\sigma$= , p\textless0.001).

\begin{figure}
  \centering
\includegraphics[width=0.95\textwidth]{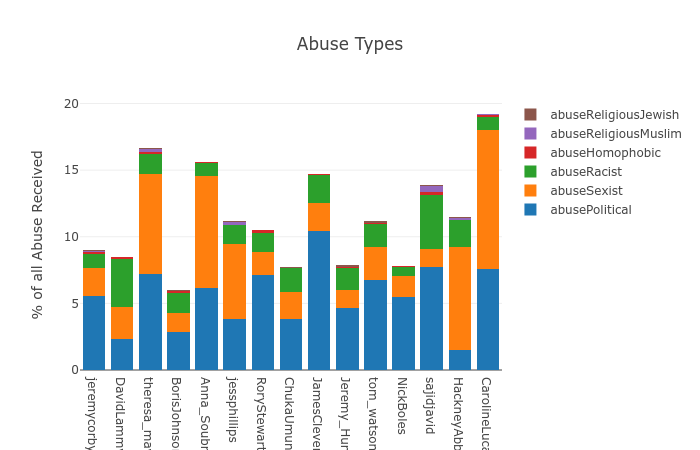}
  \caption{Abuse types received by high profile MPs}
   \label{fig:abuse-type-indiv}
\end{figure}

Figure~\ref{fig:abuse-type-indiv} shows the abuse types received by those individuals receiving the most replies in the corpus (who might be considered the highest profile MPs on Twitter). The amount of political abuse received varies a great deal, and it is interesting that the incoming prime minister, Boris Johnson, doesn't tend to attract political abuse. Diane Abbott is another notable example where the abuse she is receiving doesn't focus on her politics. The women politicians shown visibly receive more sexist abuse. The broad definition of racist abuse is perhaps evident in the fact that all the politicians shown received a certain amount of racist abuse. In some cases, the racist abuse is aimed at a third party and has been misclassified. It is visible however that Sajid Javid is particularly subjected to racist abuse, as is David Lammy, and that Mr Javid also receives Islamophobic abuse, although in fact he is not religious. There is also a significant presence of anti-white racism in the corpus.

\subsubsection{Antisemitism and Islamophobia}

Islamophobia and antisemitism are both significantly more abusive topics than is typical for the corpus across the six month period. Antisemitism received three times as much discussion as Islam, though absolute terms the amount of antisemitic abuse found in the corpus was slightly lower (194 compared with 233 items, including abuse aimed at a third party). The disproportionate discussion of antisemitism given the greater prevalence of Islamophobia in the corpus appears to arise from a large volume of criticism centred on the Labour Party and Jeremy Corbyn. Almost 80\% of the discussion of replies to MPs about antisemitism were aimed at Labour Party MPs. Actual overt antisemitic and Islamophobic hate speech in the corpus is relatively rare, but might be seen as the tip of an iceberg, in that we only detect overt, highly offensive examples targeted at MPs, and don't detect subtleties of attitude or prevalence in society outside of those politically active on Twitter. We included the terms ``zio'' and ``zionazi'' as antisemitism.

\begin{figure}
  \centering
\includegraphics[width=0.95\textwidth]{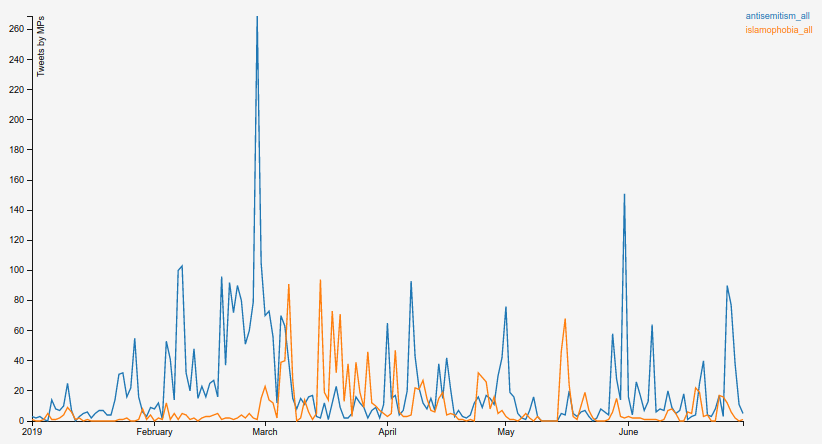}
  \caption{Antisemitism and Islamophobia discussion topics}
   \label{fig:religious-timeline}
\end{figure}

Figure~\ref{fig:religious-timeline} shows that explicit discussion of antisemitism and Islamophobia involves much lower volumes of tweets than the main topics. The greater discussion of antisemitism is clearly in evidence. This is interesting given the prevalence of topics relating to Islam in society as discussed above, and the fact that Twitter tends to be left-leaning, so we might expect more attention on the topic currently focusing criticism on the Conservative Party. On the other hand, Twitter's moderate left-of-centre majority don't tend to be as vocal as its more extreme right wing minority, and we see that Labour leader Jeremy Corbyn attracts a great deal of negative attention on Twitter, perhaps arising from this. The peak discussion of antisemitism occurred in early March. Discussion of Islamophobia mainly occurred throughout March.

\subsubsection{The ``N'' Word}

As discussed above, the racism category is broad, but nonetheless the high volumes of racist abuse are a cause for concern. The word ``n****r'' is perhaps the most offensive word in the English language, depending on context and intent, and offers an opportunity to explore usage through manual analysis. 299 examples were found of people using the word, its plural or the ``ni**a'' spelling variant, in replies to MPs or former candidates, and one example was found of an MP using that word on Twitter; David Lammy, discussed below. 185 appearances of the word were in response to Mr Lammy. Diane Abbott received 19 tweets containing the ``N'' word. Theresa May received 18 tweets containing the word. Others received fewer tweets containing the word.

However usage was extremely varied in its intent. On five occasions, tweets to Theresa May, the white outgoing prime minister, used the ``N'' word in the context of an abusive comment about people of colour. On three occasions, the ``N'' word was used as a kind of nickname for Mrs May or other readers, with no sense of offense implied. On five occasions, usage of the word is discussed, for example regarding Conservative MP Anne Marie Morris' use of the colloquial phrase ``n****r in the woodpile'' to mean undisclosed issue\footnote{\small{\url{https://en.wikipedia.org/wiki/Nigger_in_the_woodpile}}}, which caused offense. On five occasions, it was difficult to discern the intent behind the use of the word. Anne Marie Morris herself received 13 replies discussing her use of the word.

Most usage of the ``N'' word however centred on David Lammy, largely following from his late April tweet ``Words cannot express how deeply sad it is that in Britain in 2019 we have people who use the word `ni**er' running to be elected officials. This is an ideology of hate. It must be confronted and defeated.'' 131 times, people responded to his tweet with a discussion of usage of the word ``n***er''--many of these felt that he had a double standard in using the word himself yet not giving due consideration, in their opinion, to the intent behind or context of others' usage. Mr Lammy was seven times addressed as ``n***a'' with no apparent offensive intent, probably by other people of colour. On 23 occasions the word was used to abuse Mr Lammy, however. On a further nine occasions, the word featured in a reply to Mr Lammy without being used abusively; however the tweeter abused Mr Lammy using other words or slurs. On 15 occasions it was difficult to discern the intent of the tweeter.

Diane Abbott received four racist/offensive replies using the ``N'' word. She received ten replies discussing the use of the word. In one reply, the ``n***a'' variant is used as an apparently inoffensive nickname. A further four usages were difficult to classify. Sajid Javid was abused using the word on four occasions, two of which used the slur ``house n***er''. He also received a reply discussing Anne Marie Morris' use of the word. Sadiq Khan was abused using the word on four occasions, two of which were ``sand n***er''. He also received a reply in which the word was used as an inoffensive nickname.

Out of the 299 replies, 19 (6\%) used the ``N'' word as an inoffensive nickname, all bar two of those using the ``n***a'' spelling variant. 184 (62\%) discussed use of the word. 62 replies were offensive (21\%), although 12 of those don't use the ``N'' word to give offense, but feature the ``N'' word as discussion whilst using other words to give offense. This leaves 50 replies (17\%) in which the ``N'' word was used to abuse and express hatred. The remaining 44 replies (15\%) were unclear in their intent.

\subsection{RQ3: Temporal Effects in how Twitter Abuse is Experienced by MPs}

The amount of abuse an MP receives in absolute terms doesn't correspond closely to the impact it has on them. Abuse is experienced differently depending on social and personal factors, and similar abuse can be interpreted very differently depending on context. \cite{pew2017} reviewed how online abuse is experienced by American private individuals. They find a great degree of individual variation in how distressing, and how much of a problem, online abuse is perceived to be, and indeed, what is perceived as abuse and what isn't. Pew found it was more common to have been targeted for political views, with 14\% of those surveyed saying they had experienced this, than physical appearance (9\%), race (8\%), gender (8\%), religion (5\%) or sexual orientation (3\%). Men were found to be twice as likely to have been targeted for their political views, but women were twice as likely as men to have been targeted for their gender. Being targeted because of race or gender is likely be experienced as more \textit{ad hominem}, more dismissive, more damaging, than political views. Pew found that women tended to experience online abuse as more of a problem than men, which may relate to the nature of the abuse they receive.

\begin{figure}
  \centering
\includegraphics[width=0.95\textwidth]{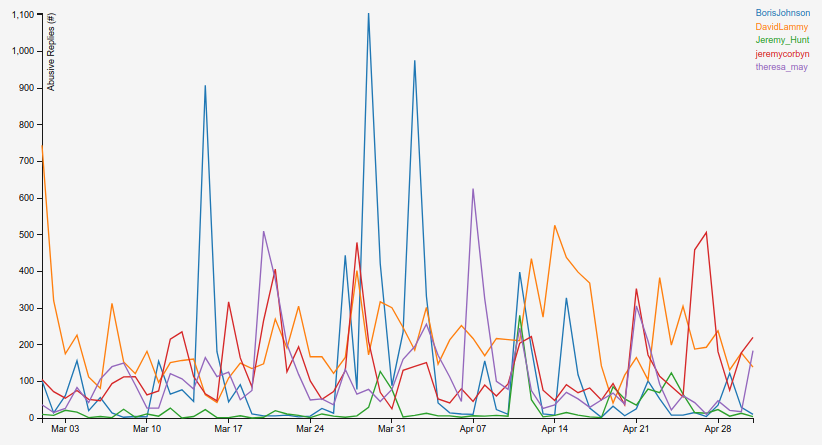}
  \caption{\% abuse over time: most abused MPs by volume}
   \label{fig:timeline-mar-apr}
\end{figure}

These findings are in keeping with those presented here. It raises the question, however, of variation in the way abuse is experienced. Temporal effects are a part of this. Averages over time and total volumes of abuse don't tend to reflect the abuse MPs feel compelled to speak out about. A steady trickle or even stream of abuse might come to be considered par for the course for a prominent public figure, whereas an intermittent deluge might feel more like bullying. Previous work by \cite{agarwal2019tweeting} has noted that Twitter attention is focused on MPs in short time windows. Figure~\ref{fig:timeline-mar-apr} is a good example of this, showing the timeline of abuse received by volume by the five most abused MPs. Timelines for January/February and May/June are available in the supplementary materials.\footnote{Withheld for anonymity.}

In this section we therefore employ the metrics used by Agarwal \textit{et al} to explore the bursty nature of online abuse; namely, focus, churn and Gini, and how that compares with the dynamics of (non-abusive) online attention.

\subsubsection{Focus}

Focus is a measure of how concentrated the abuse activity was for that individual in the time period covered by the data. Intuitively, the focus window is defined to be the longest period of consecutive days in which the MP received more than their personal average of abusive replies. Focus is then the proportion of abuse that fell within this period. So an MP who had a single burst therefore would have a high focus, whereas an MP that received more moderately varying levels of abuse would have a low focus. Formally (from Agarwal \textit{et al}), for a given MP $i$ and a threshold average abuse level $T_i$, a continuous sequence of days $R$ is a high activity window if it satisfies the property:

\begin{equation*}
    high\_activity_i(R): \sum_{d \in R}v_{id} > T_i|R|
\end{equation*}
    
Where $v_id$ is the amount of abuse received by the MP on day $d$. The focus window is the longest run. The fraction of abuse received during the focus window, $F(T)$ is therefore calculated as follows, where $V$ is the total volume of abuse received by the MP and $v_{id}$ the abuse received on that day:
    
\begin{equation*}
    F(T) = \frac{1}{V_i} \sum_{d \in T} v_{id}
\end{equation*}

Focus can be taken a step further by being normalized for the length of the focus window. Intuitively, we need to distinguish between a long window of somewhat elevated abuse vs. a short window of intense abuse, where both might give the same focus value. The proportion of abuse expected in a focus window of $|R|$ days is $|R|/D$. Focus can therefore be normalized as $F_iD/|R|$. A value of 1 would indicate a uniform abuse distribution, since the amount of abuse in the focus window matches the amount of abuse expected for that person in that duration. A high value indicates a focus window that is marked/unusual. In practice, calculating focus over a longer time period or with less sparse data (e.g. all replies as opposed to abusive replies) results in lower focus values, so the metric is sensitive to the dataset used.

Table~\ref{tab:focus} gives focus and normalized focus values for the fifteen MPs who received the most replies in absolute terms. It is noticeable, as above, that Diane Abbott had an unusual experience in this period for a prominent MP. Most prominent MPs are receiving a more consistent level of online abuse. Sajid Javid shows a similar pattern but to a lesser extent. We see that David Lammy receives a steady high level of abuse, not the steady lower level or bursty low level that are more common for high profile and lower profile MPs respectively.

No significant difference was found in focus metrics between genders, ethnic groups or the two main parties, suggesting that focus may be of interest as a metric of response to individuals and events (though that response may be a complex one depending on ethnicity among other things).

\begin{table*}
\begin{center}
\resizebox{.45\columnwidth}{!}{%
  \begin{tabular}{l|l|l}
\textbf{MP} & \textbf{Focus} & \textbf{Norm Foc}\\
\hline
Diane Abbott & 0.686 & 20.12\\
Sajid Javid & 0.269 & 9.47\\
Rory Stewart & 0.504 & 6.82\\
Jess Phillips & 0.160 & 5.62\\
Jeremy Hunt & 0.330 & 4.84\\
Caroline Lucas & 0.130 & 4.59\\
Tom Watson & 0.180 & 4.52\\
Jeremy Corbyn & 0.107 & 3.15\\
Anna Soubry & 0.204 & 2.76\\
Boris Johnson & 0.212 & 2.67\\
Chuka Umunna & 0.148 & 2.60\\
Nick Boles & 0.113 & 2.48\\
Theresa May & 0.058 & 2.05\\
David Lammy & 0.086 & 1.90\\
James Cleverly & 0.038 & 1.35\\
 \end{tabular}
 }
\caption{Focus/normalized focus for most-abused MPs}
\label{tab:focus}
\end{center}
\end{table*}

\begin{figure}
  \centering
\includegraphics[width=0.95\textwidth]{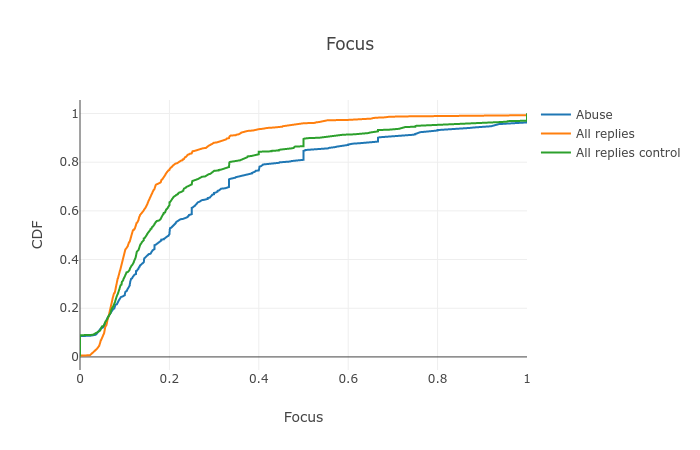}
  \caption{Focus}
   \label{fig:focus}
\end{figure}

Figure~\ref{fig:focus} shows the distribution of focus values of abuse received for all MPs. It also shows the distribution of focus values for all replies, as a comparison. Since less sparse data results in a lower focus value, to compare abuse with all replies we have plotted all replies multiplied by the proportion of abuse in the corpus, in order to be able to investigate whether abuse is burstier than all replies. The graph shows that most MPs (around 80\%) have less than half of their abuse concentrated in a focus window. Comparing the abuse curve with the corrected all replies curve, we find that abuse appears to be burstier.

\begin{figure}
  \centering
\includegraphics[width=0.95\textwidth]{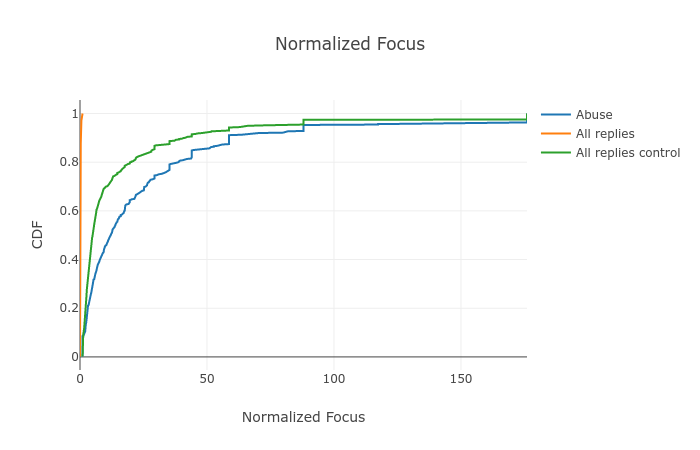}
  \caption{Normalized focus}
   \label{fig:normfocus}
\end{figure}

Figure~\ref{fig:normfocus} plots normalized focus. We find 80\% of MPs have an abuse normalized focus of 40 or less. For all replies, normalized abuse is on quite a different scale, and the short curve is comparable with that shown in \cite{agarwal2019tweeting}. However, the corrected curve again suggests that abuse is burstier, and that abuse is an exaggeration of the normal variation in social media attention received by a public figure.

\subsubsection{Churn and Gini}

Agarwal \textit{et al} also define churn as a measure of the rate of change of focus over adjacent time slices. An ``active'' MP in time slice R (which could be one or several days) is one who is currently in their focus window. Therefore:

\begin{equation*}
    Churn(R) = \frac{|active(R) \Delta active(R+)|}{|active(R) \cup active(R+)|}
\end{equation*}

\begin{figure}
  \centering
\includegraphics[width=\textwidth]{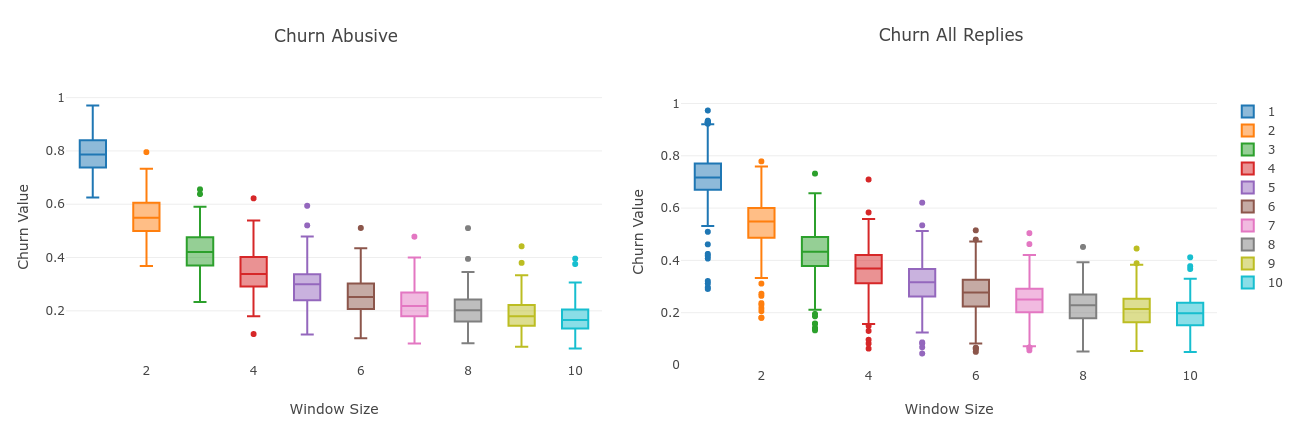}
  \caption{Churn}
   \label{fig:churn}
\end{figure}

Figure~\ref{fig:churn} gives a box plot for churn values for timeslices of one to ten days, echoing Agarwal \textit{et al}'s plot but for abusive rather than all tweets. Intuitively, we take a sliding window of $n$ days and calculate the churn of active users between the adjacent slices for each timestep. This gives a value for each timestep, the set of which are described using a box in the plot, to tell us something about that window size, before increasing $n$ and repeating. Churn isn't affected by volume of data, so there is no need to correct for this. Again we see evidence that abuse is burstier than all replies. Our work shows an interesting difference in comparison to Agarwal \textit{et al}'s. For all tweets, according to Agarwal \textit{et al}'s work, churn increases as window size increases. We, however, find churn decreases as window size increases.

\begin{figure}
  \centering
\includegraphics[width=\textwidth]{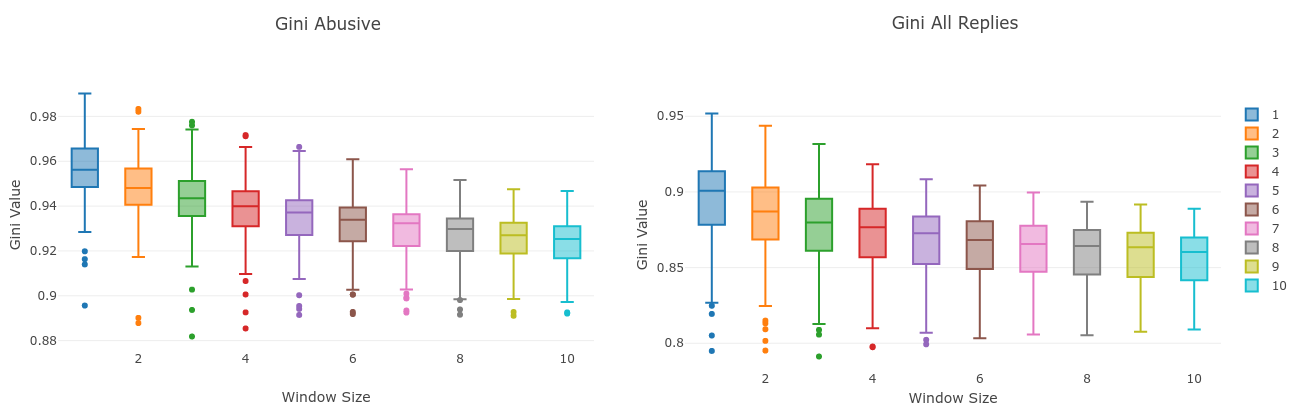}
  \caption{Gini}
   \label{fig:gini}
\end{figure}

A further metric of interest is the inequality of abuse distribution within a timeslice, as demonstrated using the well-known Gini coefficient. We have again created a box plot mirroring Agarwal \textit{et al}'s, showing different time slice durations, in figure~\ref{fig:gini}. This time, our finding of a lower Gini for larger timeslices echoes Agarwal \textit{et al}'s, and once more we find Gini is higher for abuse than for all tweets. We also find our range of Ginis remains wider for longer time windows, again demonstrating the especially sporadic and bursty nature of abusive online attention.

In summary, churn and Gini box plots, compared with Agarwal \textit{et al}'s findings, show that whilst online attention is already bursty, tending to glob together on events/individuals of interest, online abuse is even more so. We have not found evidence of race or gender effects on burstiness of online abuse, but these metrics are found to highlight the different experiences of individual politicians, and in aggregate we find that online abuse is burstier than online attention, which may affect the way it is experienced.

\section{Discussion}

Previous work about online abuse of UK MPs is continued here with a study of Q1 \& 2, 2019. Against a backdrop of tensions regarding the UK's exit from the European Union, and higher levels of online abuse toward MPs than we have previously observed, we take the opportunity to look more closely at religious and racial dynamics, as they evolve partly following from ``Brexit'' and partly as part of the global picture.

We found that several high profile ethnic minority MPs received substantial racist abuse on Twitter. David Lammy received the most abuse by volume. Diane Abbott received a significant volume of racial abuse. Sajid Javid and James Cleverly were also subjected to strong racial slurs. James Cleverly received the second highest volume of general abuse, behind David Lammy. We found that Mr Lammy received unusually high levels of abuse in response to his tweets about race. Ms Abbott received unusually high levels of abuse in response to her tweets about the repatriation of a young Muslim woman. Ethnic minorities received more racist and Islamophobic abuse and women received more sexist abuse.

In terms of more general abuse, incoming prime minister Boris Johnson attracts the highest proportion of abuse (as opposed to volume) of any high profile politician. There is a possibility of antisemitism in some of the abuse he received (Mr Johnson has a variety of heritages, including Jewish) but this is unclear. As in previous work, we found men receive more general abuse, as do Conservative politicians, possibly due to the higher proportion of men in the Conservative party, or because they are the ruling party.

Antisemitism in the Labour Party and Islamophobia in the Conservative party were topics that attracted attention during this period, echoing a sense of polarization around identity politics that is also evident in responses to David Lammy's and Diane Abbott's tweets. The topic of community and society, which includes Muslim and Jewish communities, as well as LGBT+ identities, appeared second only to Brexit, Europe and Democracy for both the Labour and Conservative parties. We found more discussion of antisemitism and more antisemitic abuse in the corpus than discussion of Islamophobia or Islamophobic abuse CHECK despite domestic events in that period being more relevant to Islam than Judaism. Recall, however, that Twitter is not a balanced sample of the populace, and within that we looked only at replies to UK MPs, so conclusions shouldn't be drawn about UK focus generally.

David Lammy's engagement with the subject of the ``N'' word during the period studied provided an opportunity to investigate in more depth responses to that word and the way it is used. The dominant use of the word was to discuss its usage. The second most frequent way of using the word is as a strong racial slur, often in conjunction with other offensive words. Significant usage also treats the word as a harmless nickname, however, especially the ``n**ga'' spelling variant, particularly ``my ni**a''.

For most MPs, abuse is bursty, more so than general online attention, which is also highly bursty. We suggest this may affect how it is perceived, as an occasional deluge may have greater shock value than a steady trickle. Burstiness of abuse varied widely between MPs, however, and is suggestive of a response to personal factors that may mean different MPs are having consistently different experiences. Diane Abbott received particularly sudden deluges compared with other prominent politicians. On the other hand, the sustained high level of abuse received by David Lammy can hardly be considered preferable. Future work is needed to explore focus and normalized focus as consistent characteristics of individuals in the public eye, as this may yield insights into what causes this difference in response and what the consequences are for politics.

\section*{Acknowledgement(s)}

This work was partially supported by the European Union under grant agreements No. 825297 ``WeVerify'' and No. 654024 ``SoBigData''.

\bibliographystyle{apacite}
\bibliography{inf-comm-soc-MP-abuse-2019}

\end{document}